\documentclass[english]{revtex4}
\usepackage[T1]{fontenc}
\usepackage[latin9]{inputenc}
\makeatletter

\providecommand{\tabularnewline}{\\}

\@ifundefined{textcolor}{}
{%
 \definecolor{BLACK}{gray}{0}
 \definecolor{WHITE}{gray}{1}
 \definecolor{RED}{rgb}{1,0,0}
 \definecolor{GREEN}{rgb}{0,1,0}
 \definecolor{BLUE}{rgb}{0,0,1}
 \definecolor{CYAN}{cmyk}{1,0,0,0}
 \definecolor{MAGENTA}{cmyk}{0,1,0,0}
 \definecolor{YELLOW}{cmyk}{0,0,1,0}
}



\makeatother

\usepackage{babel}
\begin{document}

\title{Updated values of running quark and lepton masses at GUT scale in
SM, 2HDM and MSSM}

\author{Kalpana Bora}

\affiliation{Physics Department, Gauhati University, Assam, India}

\email{kalpana@gauhati.ac.in}

\begin{abstract}
Updated values of running quark and lepton masses at GUT (Grand unified
theories) scales are important for fermion mass model building, and
to calculate neutrino masses, in GUTs . We present their values at
GUT scales, in SM, MSSM and 2HDM theories, using the latest values
of running quark and lepton masses. 
\end{abstract}
\maketitle

\section{INTRODUCTION}

Gauge theories are very attractive theories to explain the origin
of all interactions among the fundamental particles. Standard model
(SM) is a gauge theory based on group $SU(2)_{L}\times U(1)_{Y}\times SU(3)_{C}$
($G_{213}).$ In SM, all fundamental particles get their masses via
the celebrated Higgs mechanism. One of the major goals of current
research in experimental and theoretical high energy physics is to
understand the origin of all fermion masses and mixings, including
those of neutrinos. Although SM has been very successful in explaining
many of the observed experimenatl results, some questions remain unanswered
in it. Gauge hierarchy problem, unification of gauge couplings, neutrino
masses, origin of baryon asymmetry of the Universe (BAU), being the
most importane ones. Some of these problems can be circumvented if
we consider two higgs doublet model (2HDM), minimal supersymmetric
standard model (MSSM) \cite{key-1}, and GUTs. Although problem of
gauge hierarchy is not solved by 2HDM, unification of gauge couplings
is possible at GUT scales, in MSSM, and also after embedding them
in non-SUSY GUTs like SO(10) \cite{key-2}. Very recently \cite{key-3},
we have shown unification of the three gauge couplings $\alpha_{1Y}$
(for $U(1)_{Y}$), $\alpha_{2L}$ (for $SU(2)_{L}$), and $\alpha_{2C}$
(for $SU(3)_{C}$) in non-SUSY SM with additional flavor symmetries,
and also estimated limits on proton life time.

It is now a welll established fact that neutrinos have mass, and mix
with each other and oscillate to other flavors. We know that in SM,
neutrinos masses can not be explained, and hence we need to go to
theories beyond standard model (BSM). One of the most promising theories,
to explain small neutrino masses, is the grand unified theory (GUT),
like SO(10), in which all the fermions, inlcuding the right handed
(RH) neutrino, are present in a single 16-dimensional representation.
These theories require running masses and mixings of quarks and charged
leptons at GUT scales, for calculating neutrino masses. In theories
based upon quark-lepton unification, like L-R symmetric $SU(2)_{L}\times SU(2)_{R}\times SU(4)_{C}$
group, these values are also required at intermdeiate scales. Unification
of fundamental forces is based upon gauge symmetries which contain
the standard model with fermions in the fundamental representations.
Thus, the explaination of fermion masses and mixings must emerge from
a successful unified gauge theory. And hence, the running fermion
masses are required to build underlying textures and models for existence
of appropriate unified theory.

Values of running masses of quarks and charged leptons at higher scales
in SM, 2HDM and MSSM are available in literature \cite{key-4}. They
have been used quite extensively, by many researchers, e.g. in
\begin{itemize}
\item \cite{key-5}~\cite{key-6}, for constructing neutrino masses 
\item \cite{key-7}, for studying structures of unified thoeries 
\item \cite{key-8}to study type II seesaw dominance in Non-SUSY and split
SUSY SO(10) theory 
\item \cite{key-9}, for study of SO(10) models, to explain fermion masses
and mixing angles, including neutrino masses. 
\item \cite{key-10}for study of inverse seesaw in NonSUSY SO(10) theories 
\end{itemize}
But, in all these works, older values from \cite{key-4}have been
used, and new data for fermion masses are available, for using as
input at lower scales. The aim of present work is, to update these
values, and fill the gap. We have used latest data for masses and
couplings from PDG \cite{key-11}. Conversion of $\bar{MS}$ to DR
scheme is done using formulas given in \cite{key-12}, and top quark
mass is taken from \cite{key-13}. Following the anlysis of \cite{key-4},
we use RGEs for Yukawa couplings, gauge couplings and VEVs separately,
and calculate running values of fermion masses at GUT scale. These
values at other intermediate scales, calculation of neutrino masses
using them, will be presented elsewhere \cite{key-14}.

The paper has been organized as follows. In Section 2, we give a pedagogical
discussion on fermion masses. Section 3 contains methodology of, how
to run fermion masses from one energy scale to another. In Section
3, our new results, on updated values of running fermion masses, at
GUT scale $2\times10^{16}$ GeV, have been presented. Discussions
and conclusions have been given in Section 5.

\section{A Pedagogical discussion on fermion masses}

Now, we will have a pedagogical discussion on fermion masses. According
to quantum field theory (QFT), the \textquotedbl{}bare\textquotedbl{}
masses in the Lagrangian are infinite for all particles, but divergent
loop contributions to the propagator cancel them out to give finite
\textquotedbl{}dressed\textquotedbl{} masses. This is called renormalization.
These dressed particle masses are actually measured in experiments.
So in the case of an electron, for example, the experimentally measured
electron mass is an input parameter to the theory, and according to
QFT, the bare electron mass must be infinite, but the mass \textquotedbl{}runs\textquotedbl{}
from infinity at very small length scales, to a constant at very large
length scales (\textquotedbl{}IR fixed point\textquotedbl{}). So this
IR-limit value is the same as the experimentally measured value.

We know that quarks are confined, and free quarks cannot be observed
experimentally. This short distance confinement is believed to be
because of nonperturbative effects, and is associated with the scale
$\Lambda_{QCD}\sim$ 2 $GeV^{2}$. At energies greater than $\Lambda_{QCD}$$ $,
the QCD is perturbative. Since free quarks do not exist at energy
scales less than $\Lambda_{QCD}$ (also called infrared (IR) limit)
, mass for them is not well defined. Hence quark masses are scale
dependent, and they are aften defined at a energy scale. The scale
dependent quark masses are called `current' or `running' quark mass,
and they are renormalization scheme dependent. But equivalence of
these renormalization scheme-dependent quark masses can be established
with renormalization group equations (RGEs). The `constituent' quark
mass is believed to be roughly the mass that contributes to observed
mass of hadron, for example. Nonrelativistic quark models use constituent
quark masses, the constituent mass of up and down quarks are $\sim$
350 MeV.

For quarks masses also `running' takes place, but instead of converging
to a constant, they diverge at the energy scale $\Lambda_{QCD}$.
They become infinite at a much smaller length scale. This makes perfect
sense because quarks are confined into hadrons and can't be observed
macroscopically.The masses given in PDG \cite{key-11}are the values
of the `running' masses at some energy scale greater than (length
scale smaller than) $\Lambda_{QCD}$, defined in some specific renormalization
scheme.

\section{Running of masses and couplings using RGEs}

In the renormalization theories, where the Yukawa couplings and the
VEVs run separately \cite{key-15}-\cite{key-23} , the Dirac mass
of a fermion can be defined as

\begin{equation}
M_{i}(\mu)=Y_{i}(\mu)v_{i}(\mu).
\end{equation}
Here, $M_{i}(\mu)$ is the Dirac mass of the $i$-type fermion, $Y_{i}(\mu)$
is corresponding Yukawa coupling, and $v_{i}(\mu)$ is the running
VEV (Vaccum expectation value), at the scale $\mu$. In these scenarios,
the Yukawa couplings and VEVs run separately, independent of each
other. Many authors have used these \cite{key-15}-\cite{key-23},
see \cite{key-4} (Das, Parida) for a complete discussion.

The relevant terms of the Lagrangian, for masses of fermions, in SM,
can be written as:

\begin{equation}
L=\bar{q_{L}}Y_{U}\tilde{\phi}u_{R}+\bar{q_{L}}Y_{D}\phi d_{R}+\bar{l_{L}Y_{E}\phi}e_{R}+h.c.
\end{equation}
Here, $\phi$ is the higgs particle, $v(\mu)$ its running VEV at
scale $\mu$, $q_{L}$ is the left handed quark doublet, $u_{R}$
is the right handed quark, $d_{R}$ is the right handed quark, $l_{L}$
is the left handed lepton doublet, and $e_{R}$ is the right handed
electron. Since in SM, no right handed neutrinos are present, there
is no term in the Lagrangian for the neutrino mass. Similarly, for
2HDM and MSSM, this can be written as:

$ $
\begin{equation}
L=\bar{q_{L}}Y_{U}\phi_{U}u_{R}+\bar{q_{L}}Y_{D}\phi_{D}d_{R}+\bar{l_{L}Y_{E}\phi_{D}}e_{R}+h.c.
\end{equation}
Here, 
\begin{equation}
<\phi_{U}^{0}>=v_{U}(\mu)=v(\mu)sin\beta,<\phi_{D}^{0}>=v_{D}(\mu)=v(\mu)cos\beta
\end{equation}
and 
\begin{equation}
tan\beta(\mu)=v_{U}(\mu)/v_{D}(\mu)
\end{equation}
Now, we write the RGEs for running of Yukawa and gauge couplings,
for the SM, 2HDM, and MSSM, along with their RG coefficients. They
have been given in \cite{key-4}, but we present them here for the
sake of completeness only. The one-loop RGEs for Yukawa couplings,
for SM, MSSM and 2HDM, can be written as \cite{key-15}-\cite{key-18},
\cite{key-24}-\cite{key-26}
\begin{eqnarray}
16\pi^{2}\frac{dY_{U}}{dt} & = & [Tr(3Y_{U}Y_{U}^{\dagger}+3aY_{D}Y_{D}^{\dagger}+aY_{E}Y_{E}^{\dagger})\nonumber \\
 &  & +\,\frac{3}{2}(bY_{U}Y_{U}^{\dagger}+cY_{D}Y_{D}^{\dagger})-\sum_{i}C_{i}^{(u)}g_{i}^{2}]Y_{U}\\
\nonumber \\
16\pi^{2}\frac{dY_{D}}{dt} & = & [Tr(3aY_{U}Y_{U}^{\dagger}+3Y_{D}Y_{D}^{\dagger}+Y_{E}Y_{E}^{\dagger})\nonumber \\
 &  & +\,\frac{3}{2}(bY_{D}Y_{D}^{\dagger}+cY_{U}Y_{U}^{\dagger})-\sum_{i}C_{i}^{(d)}g_{i}^{2}]Y_{D}\\
\nonumber \\
16\pi^{2}\frac{dY_{E}}{dt} & = & [Tr(3aY_{U}Y_{U}^{\dagger}+3Y_{D}Y_{D}^{\dagger}+Y_{E}Y_{E}^{\dagger})\nonumber \\
 &  & +\,\frac{3}{2}bY_{E}Y_{E}^{\dagger}-\sum_{i}C_{i}^{(e)}g_{i}^{2}]Y_{E}
\end{eqnarray}
The RGEs for the VEV in SM, upto 2-loop have been derived using wave-function
renormalisation of the scalar field {[}15-16, 18-19, 21-22 {]}, and
the 1-loop equation is

\begin{equation}
16\pi^{2}\frac{dv}{dt}=[\sum_{i}C_{i}^{(v)}g_{i}^{2}-Tr(3Y_{U}Y_{U}^{\dagger}+3Y_{D}Y_{D}^{\dagger}+Y_{E}Y_{E}^{\dagger})]v
\end{equation}
Here, $t=\ln\mu$. The RGEs for $v_{a}$($a=u,d$) in the 2HDM up
to 1-loop and MSSM up to 2-loops are available in {[}15-18, 20{]}.
The 1-loop equations in both theories are

\begin{equation}
\begin{array}{cc}
16\pi^{2}\frac{dv_{u}}{dt}=[\sum_{i}C_{i}^{v}g_{i}^{2}-Tr(3Y_{U}Y_{U}^{\dagger})v_{u}\\
\\
\end{array}
\end{equation}

\begin{equation}
\begin{array}{cc}
16\pi^{2}\frac{dv_{d}}{dt}=[\sum_{i}C_{i}^{v}g_{i}^{2}-Tr(3Y_{D}Y_{D}^{\dagger}+Y_{E}Y_{E}^{\dagger})v_{u}\\
\\
\end{array}
\end{equation}
$\begin{array}{cc}
\\
\\
\end{array}$The RGE for the gauge couplings for the three models are

\begin{equation}
16\pi^{2}\frac{dg_{i}}{dt}=b_{i}g_{i}^{3}
\end{equation}
2-loop contributions are available in literature {[} 15-18, 21-26{]},
and we use them from Das, Parida\cite{key-4}.

Using above RGEs, we run the values of fermion masses, from low scale
$M_{Z}$ to higher scale $2\times10^{16}$ GeV. The input values of
running fermion masses at $M_{Z}$ have been taken from PDG \cite{key-11},
and \cite{key-12}. Our results have been presented in next section.

\section{Results}

The new results of our computations have been presented in Tables
(I-VI ). We have presented comparisons of all our results with older
values (Das, Parida, EPCJ 2001). We have used mass of the Higgs to
be 125 GeV. It can be noted that from a recent global analysis \cite{key-27},
mass of the Higgs boson has been expected to be around this value.
The scale of supersymmetry breaking, $M_{S}=1$ Tev has been used.$ $
It is worth mentioning here that some signatures of SUSY have been
detected at LHC in third family of fermions \cite{key-28}. The pole
mass of top quark is used from PDG \cite{key-11}, to be $m_{t}=172.9\pm0.6\pm0.9$
GeV. This is first converted to running mass $m_{t}(M_{Z})=172.1\pm0.6\pm0.9$,
as described in Xing et al \cite{key-12}. This value is used for
SM and 2HDM. Then, for MSSM only, we convert this running value $m_{t}(M_{Z})$
to DR(dimensional regularization) scheme value, by using Eq. (22)
of Xing et. al. \cite{key-12}, and find this to be $m_{t}(M_{Z})_{DR}=169.9\pm0.6\pm0.9$.
The latest PDG value $1/\alpha(M_{Z})=128.91$ and $\alpha_{s}(M_{Z})=0.1189\pm0.0020$
are used in our analysis. 

\begin{center}
\begin{table}

\caption{Comparison of input values at $M_{Z}$ }

\begin{centering}
\begin{tabular}{|c|c|c|}
\hline 
Fermion  & Mass (This work)  & Mass (Das,Parida)\tabularnewline
\hline 
\hline 
$m_{u}$  & $1.45{}_{-0.45}^{+0.56}$ MeV  & $2.33_{-0.45}^{+0.42}$ MeV\tabularnewline
\hline 
$m_{c}$  & $635\pm86$ MeV  & $677_{-61}^{+56}$ MeV\tabularnewline
\hline 
$m_{t}$  & $172.1\pm0.6\pm0.9$ GeV  & $181\pm13$ GeV\tabularnewline
\hline 
$m_{d}$  & $2.90{}_{-0.4}^{+0.5}$ MeV  & $4.69{}_{-0.66}^{+0.60}$ MeV\tabularnewline
\hline 
$m_{s}$  & $57.7{}_{-15.7}^{+16.8}$ MeV  & $93.4{}_{-13.0}^{+11.8}$ MeV\tabularnewline
\hline 
$m_{b}$  & $2.82_{-0.04}^{+0.09}$ GeV  & $3.0\pm0.11$ GeV\tabularnewline
\hline 
$m_{e}$  & $0.486847$ MeV  & $0.486847$ MeV\tabularnewline
\hline 
$m_{\mu}$  & $102.75138\pm0.00033$ MeV  & $102.75138\pm0.00033$ MeV\tabularnewline
\hline 
$m_{\tau}$  & $1.74669_{-0.00027}^{+0.00030}$ GeV  & $1.74669_{-0.00027}^{+0.00030}$ GeV\tabularnewline
\hline 
\end{tabular}
\par\end{centering}

\end{table}

\par\end{center}

\subsection{Running fermion masses in SM at GUT scale $=2\times10^{16}$ GeV$ $}

\begin{center}
TABLE-II COMPARISON OF FERMION MASSES IN SM, 2-LOOP 
\par\end{center}

\begin{center}
\begin{tabular}{|c|c|c|}
\hline 
Fermion  & Update Mass (This work)  & Mass (Das, Parida)\tabularnewline
\hline 
\hline 
$m_{u}$  & $0.4565_{-0.1483}^{+0.1742}$ MeV  & $0.8351{}_{-0.1700}^{+0.1636}$ MeV\tabularnewline
\hline 
$m_{c}$  & $ $$0.2225_{-0.0280}^{+0.0280}$ GeV  & $ $$0.2426{}_{-0.0247}^{+0.0235}$ GeV\tabularnewline
\hline 
$m_{t}$  & $70.5188_{-0.9479}^{+0.9585}$ GeV  & $75.4348{}_{-8.5401}^{+9.9647}$ GeV\tabularnewline
\hline 
$m_{d}$  & $ $$1.0773_{-0.4365}^{+0.4474}$ MeV  & $ $$1.7372{}_{-0.2636}^{+0.4846}$ MeV\tabularnewline
\hline 
$m_{s}$  & $20.4323_{-5.4912}^{+5.7159}$ MeV  & $34.5971{}_{-5.1971}^{+4.8857}$ MeV\tabularnewline
\hline 
$m_{b}$  & $0.9321_{-0.0172}^{+0.0166}$ GeV  & $0.9574{}_{-0.0169}^{+0.0037}$ GeV\tabularnewline
\hline 
$m_{e}$  & $0.4413\pm0.0003$ MeV  & $0.4413\pm0.0001$ MeV\tabularnewline
\hline 
$m_{\mu}$  & $93.116\mp0.0117$ MeV  & $93.1431_{-0.0101}^{+0.0136}$ MeV\tabularnewline
\hline 
$m_{\tau}$  & $1.6109\mp0.00003$ GeV  & $1.5834_{-0.0005}^{+0.000}$ GeV\tabularnewline
\hline 
\end{tabular}
\par\end{center}

\subsection{Running fermion masses in MSSM at GUT scale $=2\times10^{16}$ GeV$ $}

\begin{center}
TABLE-III COMPRISON OF MASSES IN MSSM, 2-LOOP,TAN$\beta$=10 
\par\end{center}

\begin{center}
\begin{tabular}{|c|c|c|}
\hline 
fermion  & mass(this work)  & mass (Das, Parida)\tabularnewline
\hline 
\hline 
$m_{u}$  & $0.3961_{-0.1283}^{+0.1505}$ MeV  & $0.7238{}_{-0.1467}^{+0.1365}$ MeV\tabularnewline
\hline 
$m_{c}$  & $ $$0.1930_{-0.0245}^{+0.0241}$ GeV  & $ $$0.2103{}_{-0.0212}^{+0.0190}$ GeV\tabularnewline
\hline 
$m_{t}$  & $71.0883{}_{-1.5933}^{+1.6849}$ GeV  & $82.4333{}_{-21.2264}^{+30.2676}$ GeV\tabularnewline
\hline 
$m_{d}$  & $0.9316_{-0.3769}^{+0.3858}$ MeV  & $1.5036{}_{-0.2304}^{+0.4235}$ MeV\tabularnewline
\hline 
$m_{s}$  & $ $$17.6702_{-4.6950}^{+4.9233}$ MeV  & $ $$29.9454{}_{-4.5444}^{+4.3001}$ MeV\tabularnewline
\hline 
$m_{b}$  & $0.9898_{-0.0259}^{+0.0291}$ GeV  & $1.0636{}_{-0.0865}^{+0.1414}$ GeV\tabularnewline
\hline 
$m_{e}$  & $ $$0.3585_{-0.0002}^{+0.0001}$ MeV  & $ $$0.3585\pm0.0003$ MeV\tabularnewline
\hline 
$m_{\mu}$  & $75.639\mp0.0003$ MeV  & $75.6715_{-0.0501}^{+0.0578}$ MeV\tabularnewline
\hline 
$m_{\tau}$  & $1.3146_{-0.0003}^{+0.0004}$ GeV  & $1.2922{}_{-0.0012}^{+0.0013}$ GeV\tabularnewline
\hline 
\end{tabular}
\par\end{center}

\begin{center}
TABLE-IV COMPARISON OF MASSES IN MSSM, 2-LOOP, TAN$\beta$=55 
\par\end{center}

\begin{center}
\begin{tabular}{|c|c|c|}
\hline 
fermion  & mass (this work)  & mass (Das, Parida)\tabularnewline
\hline 
\hline 
$m_{u}$  & $0.3963{}_{-0.1284}^{+0.1506}$ MeV  & $0.7244_{-0.1466}^{+0.1219}$ MeV\tabularnewline
\hline 
$m_{c}$  & $ $$0.1932{}_{-0.0246}^{+0.0240}$ GeV  & $ $$0.2105_{-0.0211}^{+0.0151}$ GeV\tabularnewline
\hline 
$m_{t}$  & $80.4472{}_{-2.6158}^{+2.9128}$ GeV  & $95.1486_{-20.6590}^{+69.2836}$ GeV\tabularnewline
\hline 
$m_{d}$  & $ $$0.9284{}_{-0.3754}^{+0.3838}$ MeV  & $ $$1.4967_{-0.2278}^{+0.4157}$ MeV\tabularnewline
\hline 
$m_{s}$  & $17.6097{}_{-4.6737}^{+4.8972}$ MeV  & $29.8135_{-4.4967}^{+4.1795}$ MeV\tabularnewline
\hline 
$m_{b}$  & $1.2424{}_{-0.0572}^{+0.0626}$ GeV  & $1.4167_{-0.1944}^{+0.4803}$ GeV\tabularnewline
\hline 
$m_{e}$  & $0.3569{}_{+0.0001}^{-0.0001}$ MeV  & $0.3565_{+0.0002}^{-0.0001}$ MeV\tabularnewline
\hline 
$m_{\mu}$  & $75.3570{}_{+0.0682}^{-0.0744}$ MeV  & $75.2938_{+0.0515}^{-0.1912}$ MeV\tabularnewline
\hline 
$m_{\tau}$  & $ $$1.6459{}_{-0.0206}^{+0.0114}$ GeV  & $ $$1.6292_{-0.0294}^{+0.0443}$ GeV\tabularnewline
\hline 
\end{tabular}
\par\end{center}

\subsection{Running fermion masses in 2HDM at GUT scale $=2\times10^{16}$ GeV$ $}

\begin{center}
TABLE-V COMPARISON OF MASSES IN 2HDM, 1-LOOP, TAN$\beta$=10 
\par\end{center}

\begin{center}
\begin{tabular}{|c|c|c|}
\hline 
fermion  & mass (this work)  & mass (Das, Parida)\tabularnewline
\hline 
\hline 
$m_{u}$  & $0.4776{}_{-}^{+}$ MeV  & $0.8749{}_{-0.1772}^{+0.1701}$ MeV\tabularnewline
\hline 
$m_{c}$  & $ $$0.2328{}_{-0.0300}^{+0.0295}$ GeV  & $ $$0.2542{}_{-0.0255}^{+0.0243}$ GeV\tabularnewline
\hline 
$m_{t}$  & $74.1053{}_{-1.10893}^{+1.1047}$ GeV  & $79.6373{}_{-9.127}^{+11.1974}$ GeV\tabularnewline
\hline 
$m_{d}$  & $ $$1.1274{}_{-0.4573}^{+0.4695}$ MeV  & $ $$1.8204{}_{-0.2743}^{+0.505}$ MeV\tabularnewline
\hline 
$m_{s}$  & $21.3821{}_{-5.7071}^{+6.0060}$ MeV  & $36.2544{}_{-5.4083}^{+5.0700}$ MeV\tabularnewline
\hline 
$m_{b}$  & $1.1615{}_{-0.0298}^{+0.0297}$ GeV  & $1.2309{}_{-0.0730}^{+0.0826}$ GeV\tabularnewline
\hline 
$m_{e}$  & $0.4407{}_{-0.0003}^{+0.0003}$ MeV  & $0.4407{}_{+0.0001}^{-0.0001}$ MeV\tabularnewline
\hline 
$m_{\mu}$  & $92.9898{}_{+0.0119}^{-0.0119}$ MeV  & $93.0197{}_{+0.0122}^{-0.0122}$ MeV\tabularnewline
\hline 
$m_{\tau}$  & $ $$1.61277{}_{+0.0003}^{-0.0002}$ GeV  & $ $$1.5851{}_{-0.0005}^{+0.0005}$ GeV\tabularnewline
\hline 
\end{tabular}
\par\end{center}

\begin{center}
TABLE-VI COMPARISON OF MASSES IN 2HDM, 1-LOOP, TAN$\beta$=55 
\par\end{center}

\begin{center}
\begin{tabular}{|c|c|c|}
\hline 
fermion  & mass (this work)  & mass (Das, Parida)\tabularnewline
\hline 
\hline 
$m_{u}$  & $0.4776{}_{-0.619}^{-0.2436}$ MeV  & $0.8749{}_{-0.1772}^{+0.1701}$ MeV\tabularnewline
\hline 
$m_{c}$  & $ $$0.2328{}_{-0.0300}^{+0.0295}$ GeV  & $ $$0.2542{}_{-0.0256}^{+0.0243}$ GeV\tabularnewline
\hline 
$m_{t}$  & $77.3752{}_{-1.4187}^{+1.4741}$ GeV  & $83.9317{}_{-10.3226}^{+13.2279}$ GeV\tabularnewline
\hline 
$m_{d}$  & $ $$1.1274{}_{-0.4573}^{+0.4695}$ MeV  & $ $$1.8204{}_{-0.2743}^{+0.505}$ MeV\tabularnewline
\hline 
$m_{s}$  & $21.3836{}_{-5.7083}^{+6.0062}$ MeV  & $36.2584{}_{-5.4099}^{+5.0720}$ MeV\tabularnewline
\hline 
$m_{b}$  & $1.3053{}_{-0.0453}^{+0.0469}$ GeV  & $1.4128{}_{-0.1162}^{+0.1353}$ GeV\tabularnewline
\hline 
$m_{e}$  & $0.4407{}_{-0.0003}^{+0.0003}$ MeV  & $0.4407{}_{+0.0001}^{-0.0001}$ MeV\tabularnewline
\hline 
$m_{\mu}$  & $93.0222{}_{+0.0111}^{-0.0111}$ MeV  & $93.0536{}_{+0.0136}^{-0.0146}$ MeV\tabularnewline
\hline 
$m_{\tau}$  & $ $$1.8138{}_{-0.0053}^{+0.0058}$ GeV  & $ $$1.7851{}_{-0.0107}^{+0.0136}$ GeV\tabularnewline
\hline 
\end{tabular}
\par\end{center}

\section{Discussions and Conclusions}

We have presented updated values of running fermion masses in SM,
2HDM and MSSM at GUT scale, at $\tan\beta=10$ and $55$, using 2-loop
RGEs for SM and MSSM, and 1-loop RGEs for the 2HDM. It can be seen
from our results (Tables I-VI ) that these new values of fermion masses
are quiet different from their older counterparts (Das, Parida \cite{key-4}).
They can be used for calculation of neutrino masses in
GUTs at higher scales, as well as for buliding of theories for fermion
mass models. Here, we would like to mention that we have verified
our calculations, by reproducing the values reported in Das, Parida
\cite{key-4}. Also, our values are different from values reported
in Xing. et. al. \cite{key-12}. This is beacuse we have used
a different scheme for running fermion masses from low scale to GUT
scale. We have used RGEs for Yukawa couplings and VEVs separately.
As discussed in text, fermion masses using this scheme \cite{key-4}have
been used extensively in literature. Hence the results presented in
this paper are very important. The values of running fermion masses
at other intermediate scales, and calculation of neutrino masses using
them will be presented elsewhere \cite{key-14}.

\section{Acknowledgements}

The author would like to thank M.K. Parida for fruitful discussions
and C. R. Das for e-discussion. She also thanks UGC-SAP programme
of Physics Department, Gauhati University, India and Neutrino Project
of HRI, Allahabad, India, for providing financial assistance to visit
HRI, where some parts of the work were carried out.

\end{document}